\documentclass[aps,pre,twocolumn,longbibliography]{revtex4-2}
\usepackage{amsmath}
\usepackage{graphicx}
\usepackage{dsfont}
\usepackage{subfigure}
\usepackage{natbib}
\usepackage{siunitx}
\usepackage{xcolor}

\begin{document}
	
	
	\title{Statistical Inference of 1D Persistent Nonlinear Time Series and Application to Predictions}
	
	
	\author{Johannes A. Kassel}
	\email[]{jkassel@pks.mpg.de}
	\affiliation{Max Planck Institute for the Physics of Complex Systems, N\"othnitzer Stra{\ss}e 38, 01187 Dresden, Germany, EU}
	\author{Holger Kantz}
	\email[]{kantz@pks.mpg.de}
	\affiliation{Max Planck Institute for the Physics of Complex Systems, N\"othnitzer Stra{\ss}e 38, 01187 Dresden, Germany, EU}
	
	\date{\today}
	
	\begin{abstract}
		We introduce a method for reconstructing macroscopic models of
		one-dimensional stochastic processes with long-range correlations from
		sparsely sampled time series by combining fractional calculus and
		discrete-time Langevin equations. The method is illustrated for the
		ARFIMA(1,d,0) process and a nonlinear autoregressive toy model with multiplicative noise.
		We reconstruct a model for daily mean temperature data recorded at Potsdam (Germany) and use it to predict
		the first frost date by computing the mean first passage time of the
		reconstructed process and the $0\;\si{\celsius}$ temperature line, illustrating the potential of long-memory models for predictions in the subseasonal-to-seasonal range.
	\end{abstract}
	
	
	\maketitle
	
	\section{Introduction}
\begin{figure}
		\centering
		\includegraphics[width=.45\textwidth]{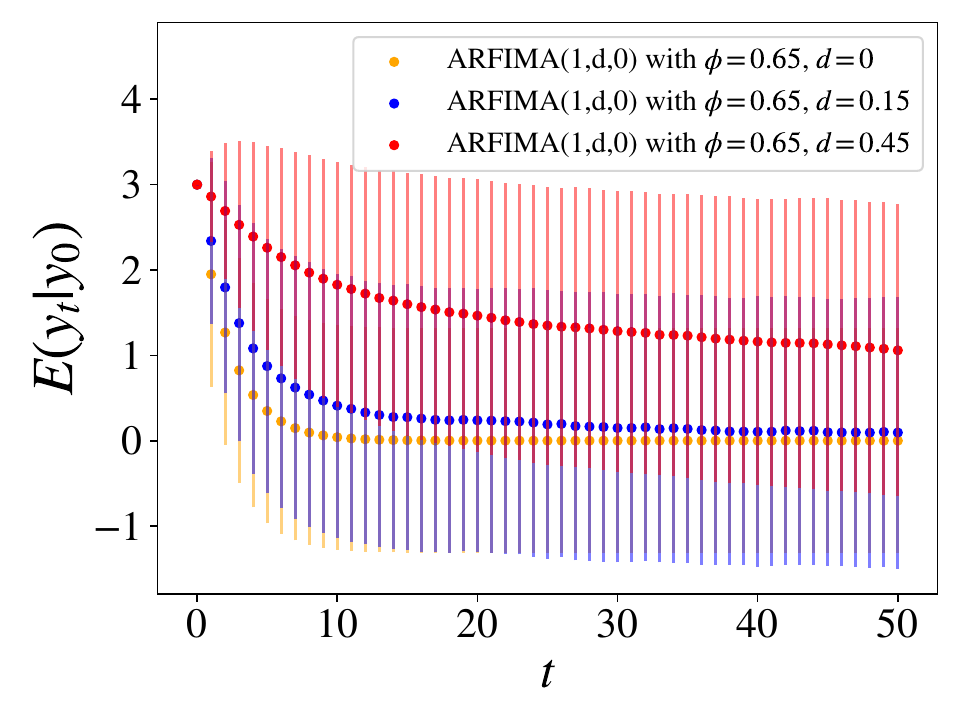}
		\caption{Conditional averages of ARFIMA(1,d,0) processes with
			$\phi=0.65$ and parameter values $d\in\{0.0, 0.15, 0.45\}$ relax to zero on different time scales. For $d=0.0$, the process simplifies to the Markovian $AR(1)$ process (yellow curve -- analytical). The displayed curves for $d\neq0$ are ensemble averages with $N=10^4$ independent samples, conditioned on $y_0 \in [2.9995, 3.0005]$ and a finite memory length $M=250$, truncating the noise integration (cf. Eq.~\ref{eq:arfima1d0}).
			Error bars indicate standard deviations. For larger $d$, the memory of the noise is stronger, resulting in a slower relaxation towards the mean of the process.
			This indicates that for processes with long-range correlations
			($d>0$), prediction horizons are longer than for processes
			without long-range correlations.}
		\label{fig:ar1_arfima}
	\end{figure}

	Predicting the dynamics of complex systems with
	models inferred from data has been a long-standing endeavor of
	science. If such models are stochastic they can capture quite naturally
	erratic fluctuations in the observed data. We will discuss the large body
	of literature on the reconstruction of Markov processes below. However,
	in many real world data sets, violations of Markovianity by long-range
	temporal correlations have been observed.
	For a stationary process with light-tailed increment distribution,
	the Hurst exponent $H$ measures such temporal long-range
	correlations \cite{Mandelbrot1968noah, Chen2017}.
	For $H>0.5$, the process exhibits persistent long-range
	correlations. For short-range correlated processes, in particular Markov processes, there exists a characteristic time scale, i.e. a minimal time separation required between two states of the process to be considered independent. Hence the process possesses no asymptotic self-similarity, resulting in $H=0.5$ \cite{Mandelbrot1968, Mandelbrot1969, Watkins2019}.
	Models for long-range correlations emerged after Hurst's
	study of the reservoir capacity for the river Nile \cite{Hurst1951}.
	Later on, long-range correlations were found in data sets of temperature
	anomalies \cite{Fraedrich2003,Eichner2003}, river runoffs
	\cite{Kantelhardt2006}, extreme events return intervals
	\cite{Bunde2005}, biological systems \cite{Wan2014, Echeverria2003}, and economics \cite{Baillie1995}.
	The earliest models generating long-range correlations are Fractional Brownian
	Motion (FBM) \cite{Mandelbrot1968} in continuous time and autoregressive fractionally integrated moving average (ARFIMA)
	processes \cite{Granger1980,Hosking1981} in discrete time.
	The ARFIMA(1,d,0) process is defined as:
	\begin{align}
	y_{t+1} &= \phi \, y_t + (1 - B)^{-d} \, \xi_t \nonumber\\
			&= \phi \, y_t +  \lim_{M\to\infty} \sum_{j=0}^{M} \frac{\Gamma(j + d)}{\Gamma(j+1)\,\Gamma(d)} \, \xi_{t-j}\,, \label{eq:arfima1d0}
	\end{align}
	in which the positive real number $\phi$ is the autoregressive
	parameter, $B$ the backshift operator, $\Gamma$ the gamma function,
	and $\xi_t$ Gaussian white noise. It has the asymptotic Hurst exponent $H=0.5 + d$
	and as Eq.~\ref{eq:arfima1d0} shows explicitly, it is not Markovian.
	Figure~\ref{fig:ar1_arfima} shows conditional averages of $y_t$,
	$E(y_t|y_0\in [2.9995, 3.0005])$ as a
	function of $t$ for various values of the memory parameter $d$, where
	the condition requires that $y_0\in [2.9995, 3.0005]$, and $E(\cdot)$ denotes
	the expectation value.
	The short-range limit of this example, $d=0$, $H=1/2$, is the Markovian
	$AR(1)$ process and has an autocorrelation	time of
	$\tau = - 1/\ln \phi \approx 2.3$. The much faster relaxation of this
	conditional mean to the sample mean of the process
	(which is 0) demonstrates that memory in the noise can lead to
	enhanced predictability of the process. Therefore, it is beneficial to
	reconstruct such models from data, if there are clear indications for
	temporal long-range correlations, instead of ignoring them.

	Today, there are many approaches to reconstructing stochastic models from
	data. Examples include Generalized Langevin equations
	\cite{Kou2004,Lei2016}, Fractional Klein-Kramers equations
	\cite{Dieterich2008}, underdamped Langevin equations
	\cite{Brueckner2020}, Fokker-Planck equations
	\cite{Honisch2012,Ragwitz2001,Boettcher2006,Tabar2019}, and discrete-time
	ARFIMA \cite{Graves2015} and nonlinear autoregressive moving average
	(NARMA) \cite{Chorin2015} models.
	While all of these approaches deal with
	either low sampling rates, long-range correlated data, nonlinear drift
	terms, multiplicative noise or single-trajectory data, none of them
	covers all of these complications for model reconstruction at
	once. However, in many applications e.g. geophysical time series recordings,
	neither trajectory ensembles nor highly sampled data sets are available,
	when the time series exhibit both non-trivial short-range and long-range behavior.
	Kir\'aly and J\'anosi propose a method for the model reconstruction of daily
	temperature anomalies with long-range correlated input noise in an ad hoc and
	approximate way. \cite{Janosi2002}
	Here, we extend this pioneering work to a generally valid framework for the
	reconstruction of discrete-time models and illustrate the predicitive
	power of long-memory models.
	
	In the remainder of this article, we describe our method and illustrate
	it by applying it to the ARFIMA(1,d,0) process, to a non-trivial toy model,
	and to daily mean temperature data. Finally, we use a reconstructed stochastic
	model of daily mean temperature anomalies to predict the first frost date
	in Potsdam, Germany, and assess the performance of the prediction.

	\section{Method}
	We exploit the scale freedom
	of long-range correlations and decompose the long-range and
	short-range behavior of stochastic time series. Firstly, we remove
	long-range correlations using the Gr\"unwald-Letnikov fractional
	derivative resulting in a process which is approximately
	Markovian. Then, we reconstruct the short-range dynamics with a dicrete-time Langevin equation. Finally, we numerically create sample paths with the inferred Langevin equation and introduce long-range temporal correlations again employing the Gr\"unwald-Letnikov fractional integral also used in ARFIMA processes.
	
	We start with a
	one-dimensional, stationary time series $\{y_t\}_{1\leq t \leq N}$ of length $N$,
	which exhibits an asymptotically constant Hurst exponent $H > 0.5$.
	The numerical value of $H$ may be determined by Detrended Fluctuation Analysis (DFA) \cite{Peng1994,Hoell2019} or other methods, among them R/S statistics \cite{Hurst1951}, and
	Wavelet transforms \cite{Simonsen1998,Abry1998}.
	We use the first-order finite difference approximation of the
	Gr\"unwald-Letnikov fractional derivative of order $d~=~H-\frac{1}{2}$ with a finite difference of $\Delta t = 1.0$, defined as \cite{Podlubny1998}
	\begin{align}
	{}_{t-M}D^{d}_{t} \; y_t = \sum_{j=0}^{M} \omega_j^{(d)} \, y_{t-j} ;\quad\label{eq:GL_derivative} \omega_j^{(d)}=(-1)^j {d\choose j}.
	\end{align}
	Here, $M$ defines the 
	memory length of the fractional operation. In theory, $M$ goes to infinity
	for fractional processes (cf. Eq.~\ref{eq:arfima1d0}). In applications,
	choosing an appropriate finite $M$ is a trade-off between the loss of $M$
	data points and the time scale of the long-range correlations to be
	removed. Choosing $M = N/2$ would be optimal, but increased
	statistical fluctuations in the subsequent analysis advice smaller
	$M$. Removal of long-range correlations from time series using
	fractional calculus has been applied e.g. in
	\cite{Petras2019,Yuan2013}. For numerical ease we use the recurrence
	relation $w^{(d)}_j =(1 - \frac{d + 1}{j}) \, w^{(d)}_{j-1}$ with
	$w_0^{(d)} = 1$ for the computation of the coefficients in
	Eq.~\ref{eq:GL_derivative}.

	The values of the resulting fractionally differenced time series are
	denoted by $\{{}_{t-M}D^{d}_{t} \, y_t \} = \{x_t\}$, which we
	consider approximately Markovian.
	We now model the time series $\{x_t\}_{1<t<N-M}$ with a stochastic difference equation \cite{Tong1993} and call it discrete-time Langevin equation
	\begin{align}
	x_{t+1} = f(x_t) + g(x_t) \, \xi_t \label{eq:discrete_langevin} \,.
	\end{align}
	Reminiscent of the continuous-time Langevin equation we refer to $f(x_t)$ as drift and to $g(x_t)$ as diffusion. Here, both $f(x_t)$ and $g(x_t)$ are allowed to be nonlinear resulting in a nonlinear restoring force and multiplicative noise, $\xi_t$ denotes Gaussian white noise with $\langle \xi_t \rangle = 0$ and $\langle \xi_t \xi_{t'} \rangle = \delta_{tt'}$. We assume $g(x_t)\geq 0$ for $x_t\in(-\infty,\infty)$. The subsequent scheme is inspired by the reconstruction scheme for time-discrete NARMA models \cite{Chorin2015, Lu2016}. At first, we make an ansatz $\Phi(x_t; \lambda), \, \lambda=(\lambda_1, \lambda_2, ...)$ for the drift $f(x_t)$. The functional form of $\Phi$ requires an educated guess upon inspection of the data in the $(x_{t+1}, x_t)$ plane. Demanding stability of the process requires $f(x_t)$ to monotonically decrease in $x_t$ for $x_t\rightarrow\pm\infty$. We then find the optimal parameters $\hat{\lambda}$ by a least-squares fit, i.e.
	\begin{align}
	\hat{\lambda} = \underset{\{\lambda\}}{\mathrm{arg}\min} \sum_{t=1}^{N-1} (x_{t+1} - \Phi(x_t; \lambda))^2 =  \underset{\{\lambda\}}{\mathrm{arg}\min} \sum_{t=1}^{N-1} {R_t(\lambda)}^2 \label{eq:residuals}\,.
	\end{align}
	For a drift function $\Phi(x_t, \hat{\lambda})$ which resembles $f(x_t)$, the averaged squared residual amounts to $\langle R_t^2 \rangle = g(x_t)^2 \langle \xi_t^2 \rangle  = g(x_t)^2$, because of assumptions about the noise. Hence, we make an ansatz $\Theta(x_t;\theta), \, \theta=(\theta_1, \theta_2,...)$ for the squared residuals. Again, an educated guess is needed for its functional form. Performing a least-squares fit yields the optimal parameters for approximating $g(x_t)^2$.
	
	With the acquired parameters, we can generate trajectories employing the following discrete-time Langevin equation:
	\begin{align}
	x_{t+1} = \Phi(x_t, \hat{\lambda}) + \sqrt{\Theta(x_t, \hat{\theta})} \; \xi_t \,.\label{eq:discrete_langevin_model}
	\end{align}
	Here, $\xi_t$ is Gaussian white noise with zero mean and variance one.
	By construction, time series generated using Eq.~\ref{eq:discrete_langevin_model} are Markovian and should have similar stochastic properties as the fractionally differenced time series
	$\{x_t\}$.
	
	Finally, we fractionally integrate the model time series,
	adding long-range correlations to the model data. For this purpose,
	we employ the first-order finite difference approximation of the
	Gr\"unwald-Letnikov fractional integral which is obtained by setting
	$d\rightarrow-d$ in Eq.~\ref{eq:GL_derivative} and reads:
	\begin{align}
	{}_{t-M}I^{d}_{t} \; x_t = \sum_{j=0}^{M} (-1)^j {-d\choose j} \, x_{t-j} \,.\label{eq:GL_integral}
	\end{align}
	Our approach neglects measurement noise. Since we are interested in
	reconstructing a macroscopic model possessing the same statistical
	properties as the original time series, we consider potential measurement
	noise as an indistinguishable part of the process.
	Choosing appropriate functions $\Phi$ and $\Theta$ is crucial for obtaining a suitable model. Therefore, we advise testing various functions and base the selection both on goodness of fit as well as comparisons of model data and original data. The assumed Markovianity of the fractionally differenced data should be tested in applications. If it is not satisfied, the discrete-time Langevin equation presented here must be replaced by a higher-order Markovian model incorporating more than one previous realization of the process.
	
	\section{ARFIMA(1,d,0) process and the discrete-time Langevin equation} We demonstrate the two parts of our method with the ARFIMA(1,d,0) process and a toy model defined by a non-linear discrete-time Langevin equation.
	From the definition of the ARFIMA(1,d,0) process $y_t$ (cf. Eq.~\ref{eq:arfima1d0}), it is clear that by applying the finite difference fractional derivative (cf. Eq.~\ref{eq:GL_derivative}) we obtain the AR(1) process:
	\begin{align*}
	x_{t+1} = \phi \, x_t + \xi_t \,,\quad 	x_t = (1 - B)^{d} \, y_t = \lim\limits_{M\rightarrow\infty} {}_{t-M}D^{d}_{t} \; y_t \,.
	\end{align*}
	Due to linearity, the autoregressive parameter $\phi$ is the same as in the ARFIMA(1,d,0) model. Hence, inference of $\phi$ from the fractionally differenced process and subsequent fractional integration of the inferred process yields the original process here. 
	
\begin{figure}
		\centering
		\subfigure[Drift of bimodal toy model\label{subfig:toy_model_drift}]{\includegraphics[width=0.235\textwidth]{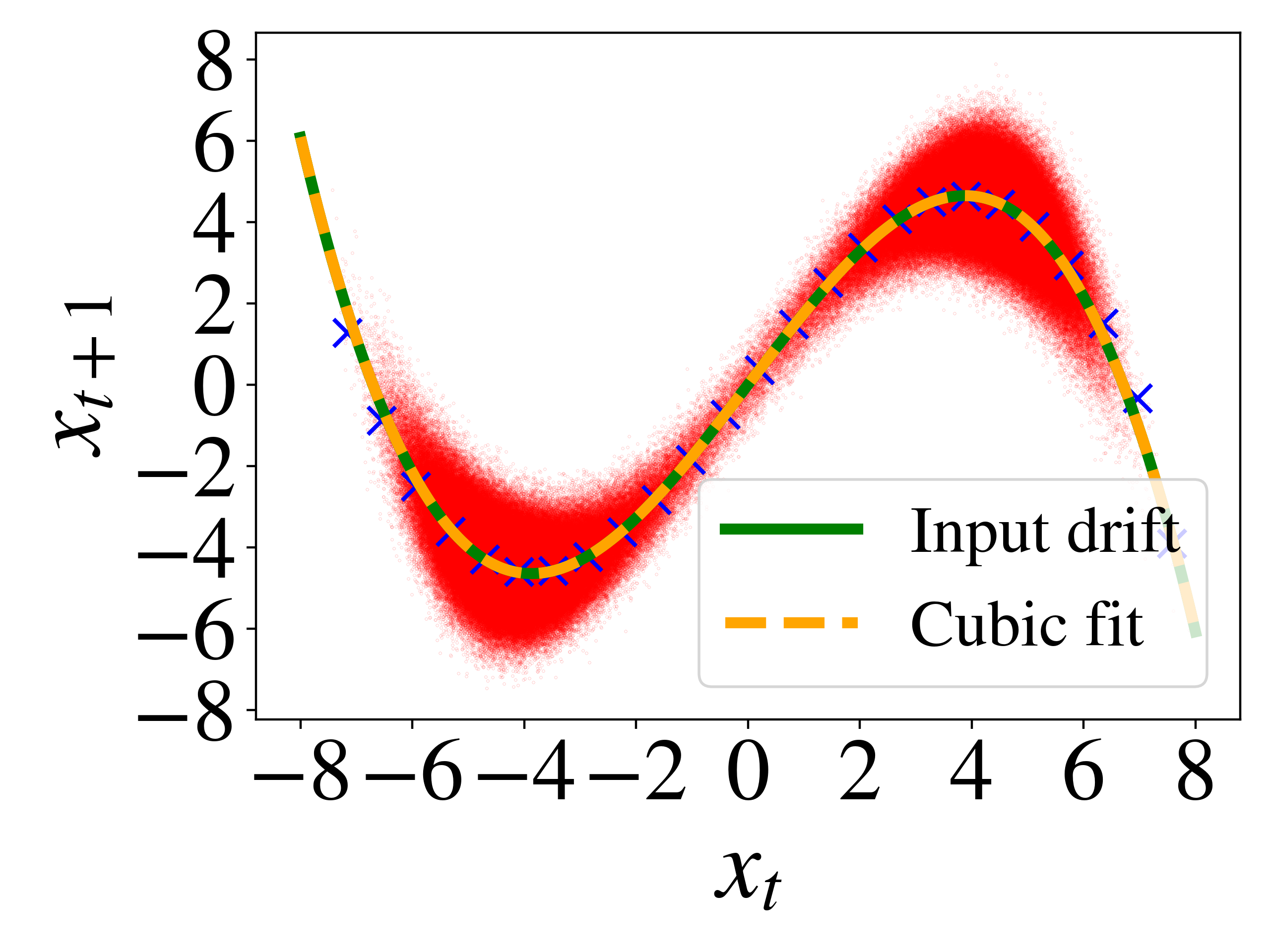}}
		\subfigure[Diffusion of bimodal toy model\label{subfig:toy_model_diff}]{\includegraphics[width=0.235\textwidth]{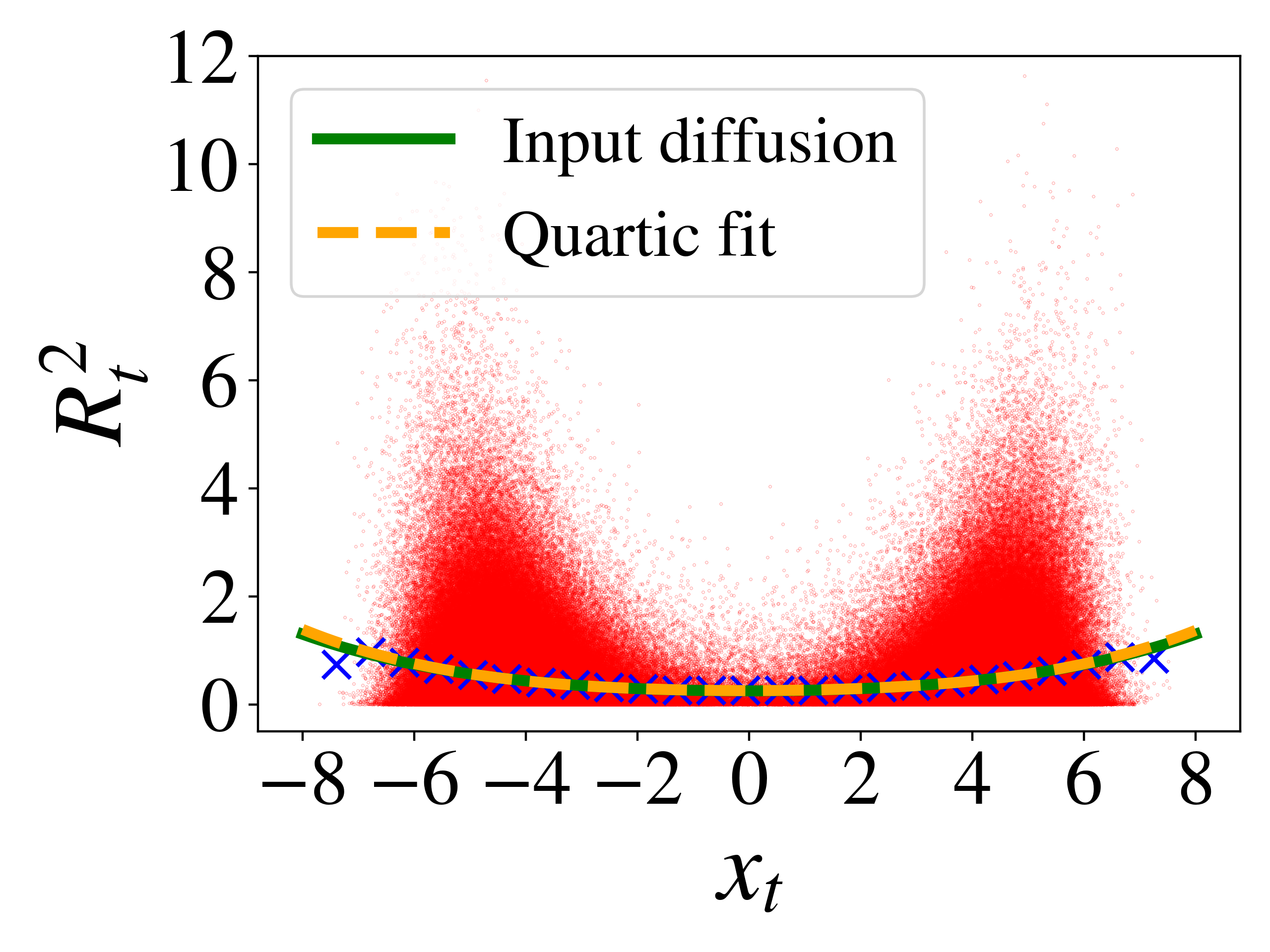}}
		\caption{Parameter inference for toy model defined by Eq.~\ref{eq:toy_model}. Left panel~\ref{subfig:toy_model_drift} shows the drift inference of the model, right panel~\ref{subfig:toy_model_diff} shows the diffusion inference of the model. Red dots are the $N=10^6$ data points. Blue crosses show average values for $25$ bins of equal width, only shown for illustration. Orange curves show the  results of least-squares fits for polynomials of order three, and four, respectively. Green dashed curves show input drift and input diffusion, respectively. Orange and green curves are in perfect agreement.}
		\label{fig:toy_model}
	\end{figure}

	The following toy model process possesses a bimodal distribution and illustrates solely the second part of our method for nonlinear functions $f(x_t)$ and $g(x_t)$:
	\begin{align}
	x_{t+1} &= - 0.04 \, x_t^3 + 1.8 \, x_t + (0.01 \, x_t^2 + 0.5) \, \xi_t \,,\label{eq:toy_model}
	\end{align}
	with $\xi_t$ as before. We make polynomial ansatzes of order three and four for the drift $\Phi(x_t)$ and diffusion $\Theta(x_t)$, respectively. Figure~\ref{fig:toy_model} displays model data as well as the perfect agreement of input drift and diffusion functions and their reconstructions. The reconstruction works also with a fifth order polynomial for $\Phi(x_t)$ and a sixth order polynomial for $\Theta(x_t)$.

	\section{Daily Temperature Data and First Frost Prediction} 
\begin{figure}
		\centering
		\includegraphics[width=.45\textwidth]{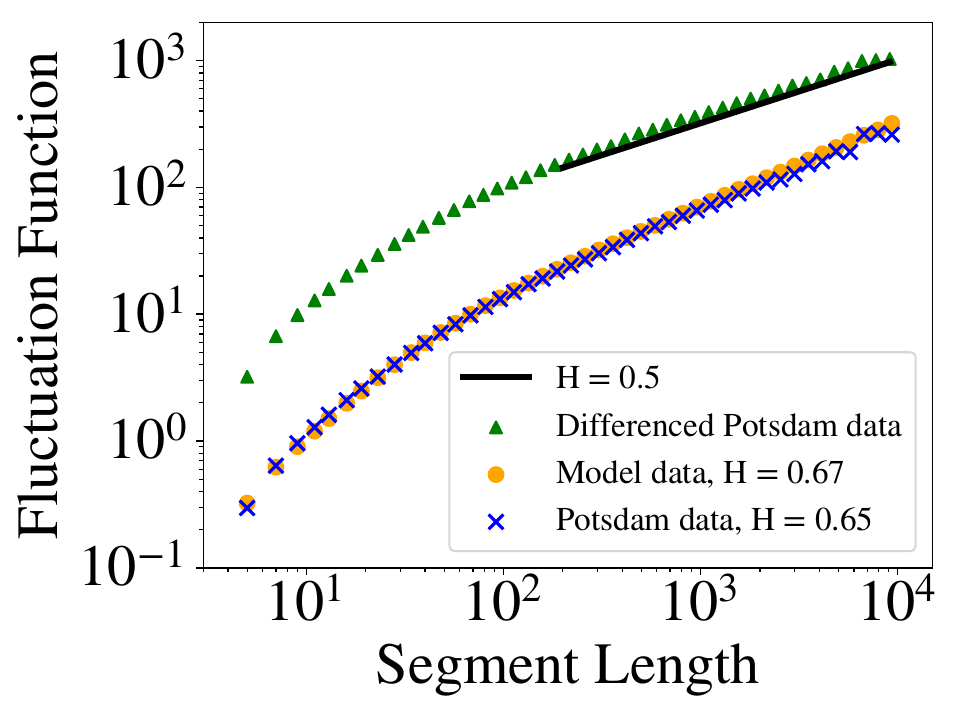}
		\caption{Detrended Fluctuation Analysis (DFA-3) of daily mean temperature anomalies (green triangles),  fractionally differenced daily mean temperature anomalies (blue crosses) and model data (orange dots). Offset for improved visibility. The asymptotic slope of the fluctuation functions $H$ of the daily mean temperature anomalies and the model data coincide almost perfectly. The slope of the fractionally differenced daily temperature anomalies approaches the $H=0.5$ line, indicating the absence of long-range correlations.}
		\label{fig:dfa_anomalies}
	\end{figure}

We apply
	our method to daily mean 2m-temperature data of the Potsdam
	Telegrafenberg weather station and predict the first frost date in
	late autumn using the first passage time of the reconstructed process
	with the zero temperature boundary. The data is provided by the
	European Climate Assessment \& Dataset project team
	(\emph{https://www.ecad.eu/}) \cite{ECAD}. The Potsdam temperature
	data set consists of an uninterrupted time series starting January 1st
	1893 and is therefore apt for our analysis. Neglecting the daily
	temperature cycle, we consider the temperature data set as a time series of a discrete-time
	stochastic process with two additional trends, namely seasonal cycle (also called climatology)
	and climate change. We approximate the seasonal cycle by fitting a second-order Fourier
	series to the data, adding a quadratic function in
	time to account for the nonstationarity of the temperature time
	series due to climate change. The resulting stationary time series
	referred to as temperature anomalies is approximately
	Gaussian \cite[Fig.2, p.9246]{Massah2016}. Here, we use DFA-3, in which a cubic polynomial is used for the detrending procedure \cite{Hoell2019}, to determine
	the Hurst exponent resulting in $H=0.65$ (cf.~Figure~\ref{fig:dfa_anomalies}).

	Following the recipe described above, we fractionally
	differentiate the temperature anomalies with $d=H-0.5$ and a memory
	length of three years ($M=1095$). Choosing longer memory ranges
	does not improve the model. The approximate Markovianity of the
	fractionally differenced data is indicated by its Hurst exponent (cf. Fig.~\ref{fig:dfa_anomalies}), the exponential decay of its autocorrelation
	function (cf. Figure~\ref{fig:acf_full_model}), and an inspection of the
	dependence of the residuals $R_t(X_{t-2}, X_{t-3}|\hat{\lambda})$ on
	previous realizations of the process, which is negligible.
	
\begin{figure}
		\centering
		\subfigure[Drift estimation\label{subfig:potsdam_drift}]{
			\includegraphics[width=.225\textwidth]{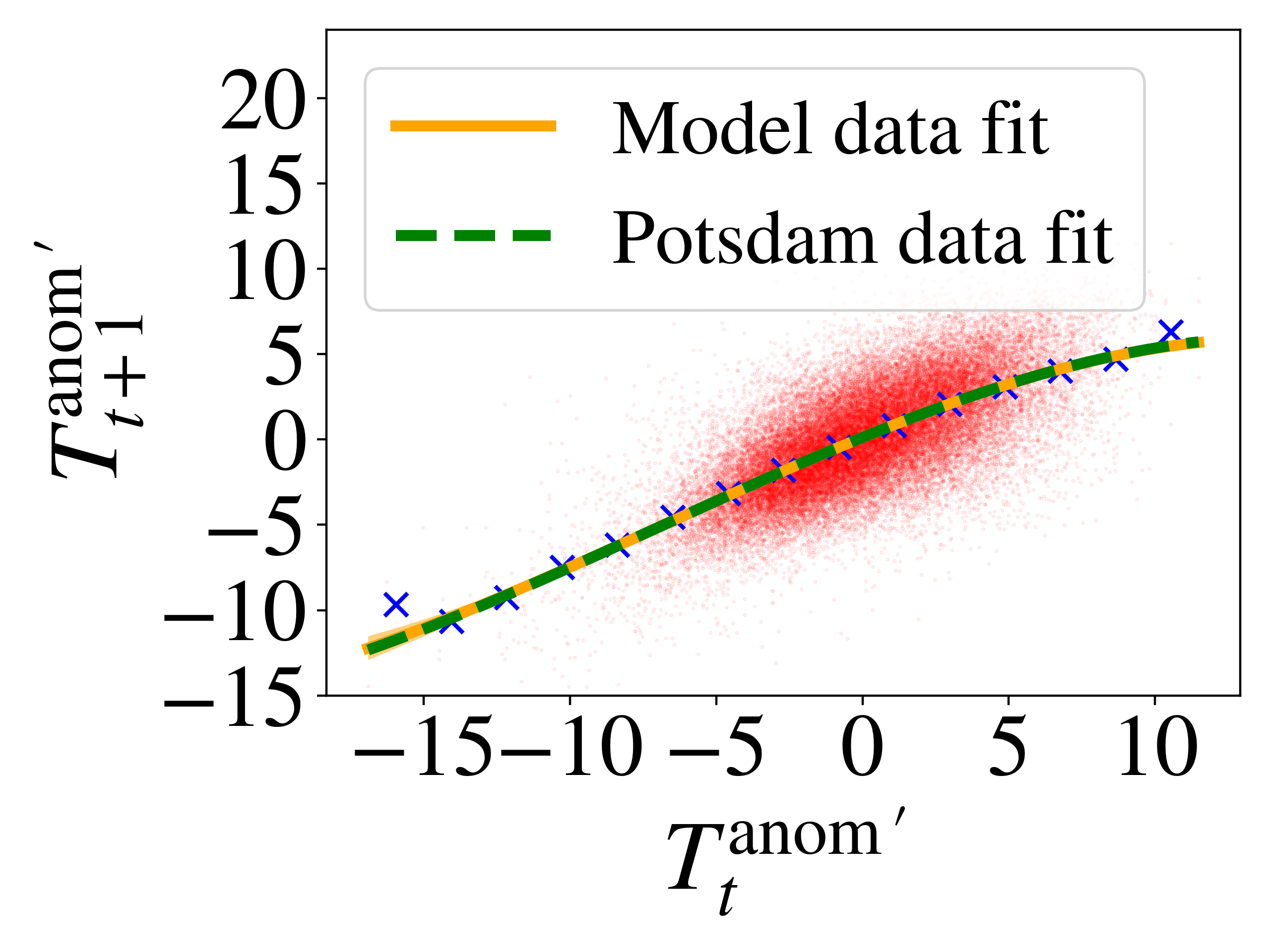}
		}
		\subfigure[Diffusion estimation\label{subfig:potsdam_diffusion}]{
			\includegraphics[width=.225\textwidth]{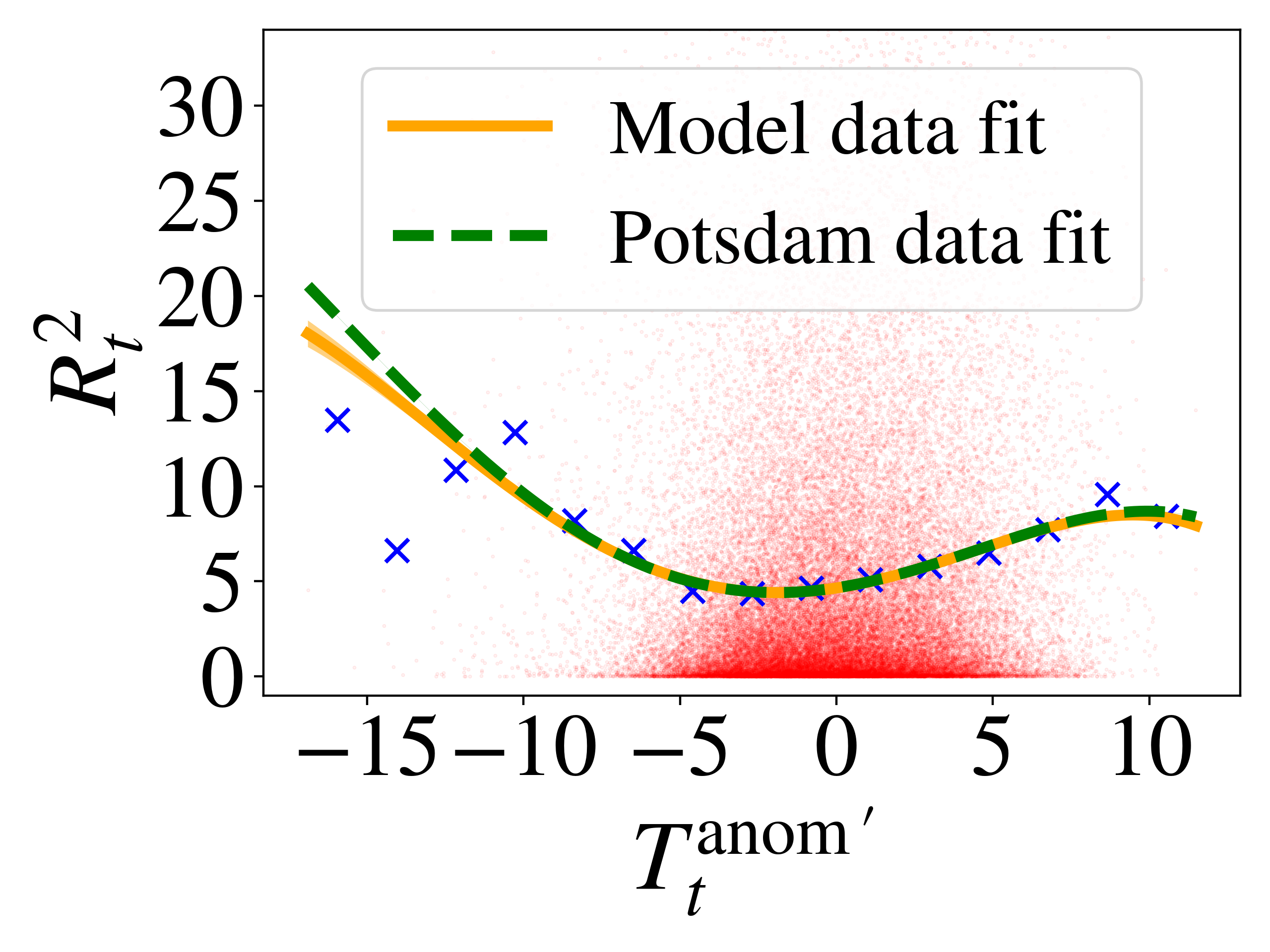}
		}
		\caption{Estimation of drift (cf. panel~\ref{subfig:potsdam_drift}) and diffusion (cf. panel~\ref{subfig:potsdam_diffusion}) of the discrete-time Langevin equation for fractionally differenced daily mean temperature anomalies of the Potsdam Telegrafenberg weather station. Red dots are the fractionally differenced anomalies (see panel~\ref{subfig:potsdam_drift}), and their squared residuals (cf. Eq.~\ref{eq:residuals}, panel~\ref{subfig:potsdam_diffusion}). The blue crosses are bin averages of the red dots, displayed for illustration only. The green curves are results of least-squares fits of polynomials of order three for the drift and order four for the diffusion. The orange curves are results of least-square fits of model data ($100$ samples of the length of the Potsdam data) generated with Eq.~\ref{eq:discrete_langevin_model} and obtained parameters of the green curves. There are small deviations of the diffusion for large negative anomalies between the Potsdam data and the modal data due to the numerical stability constraint.}
		\label{fig:KMA}
	\end{figure}

\begin{figure}
		\centering
		\subfigure[Autocorrelation function\label{fig:acf_full_model}]{
			\includegraphics[width=.45\textwidth]{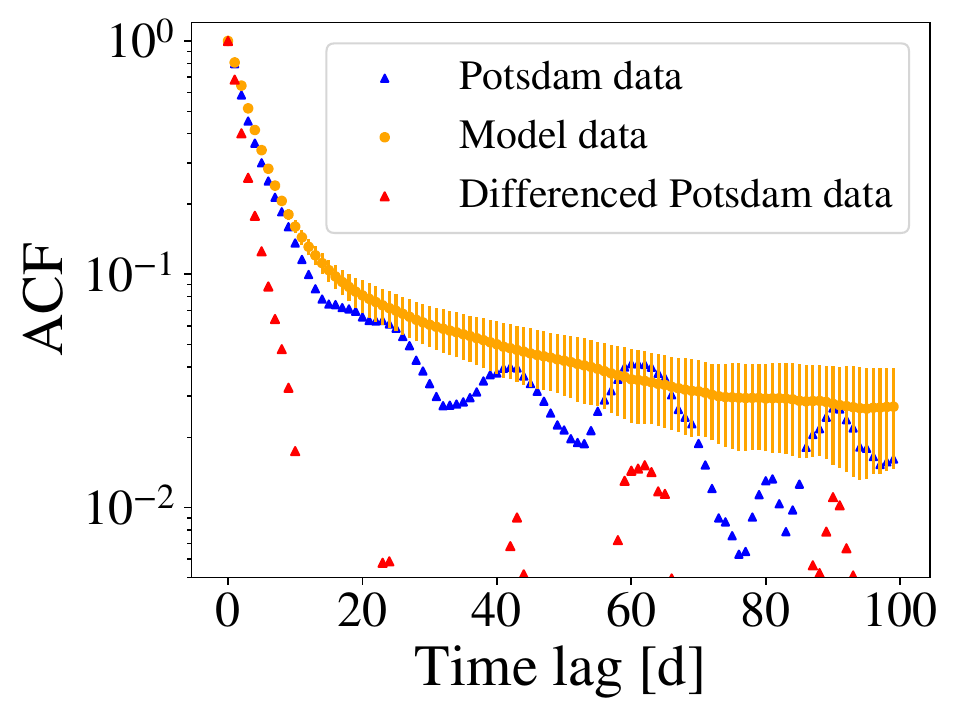}}
		\subfigure[Cumulative histogram\label{fig:hist_full_model}]{
			\includegraphics[width=.23\textwidth]{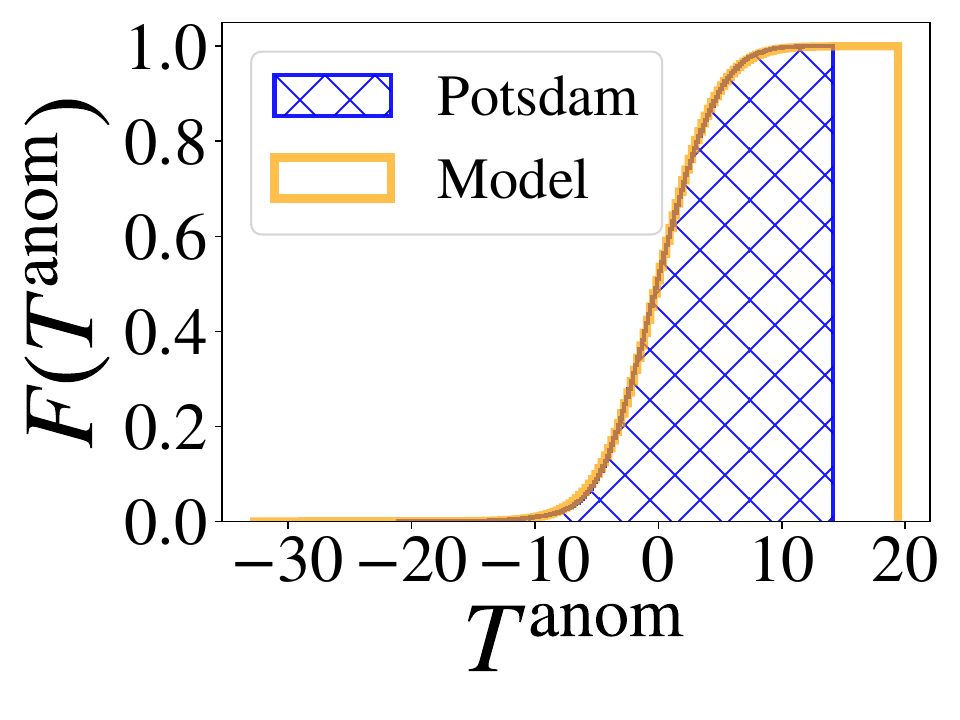}}
		\subfigure[Power spectral density\label{fig:psd_full_model}]{\includegraphics[width=0.23\textwidth]{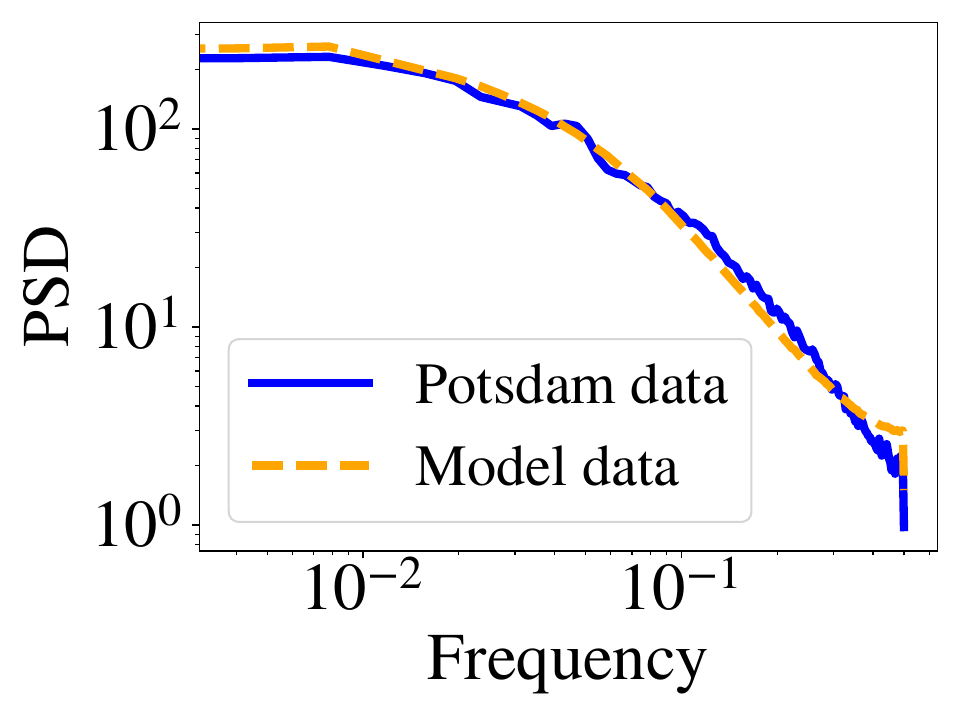}}
		\subfigure[Sample trajectory\label{fig:sample_traj_full_model}]{\includegraphics[width=0.42\textwidth]{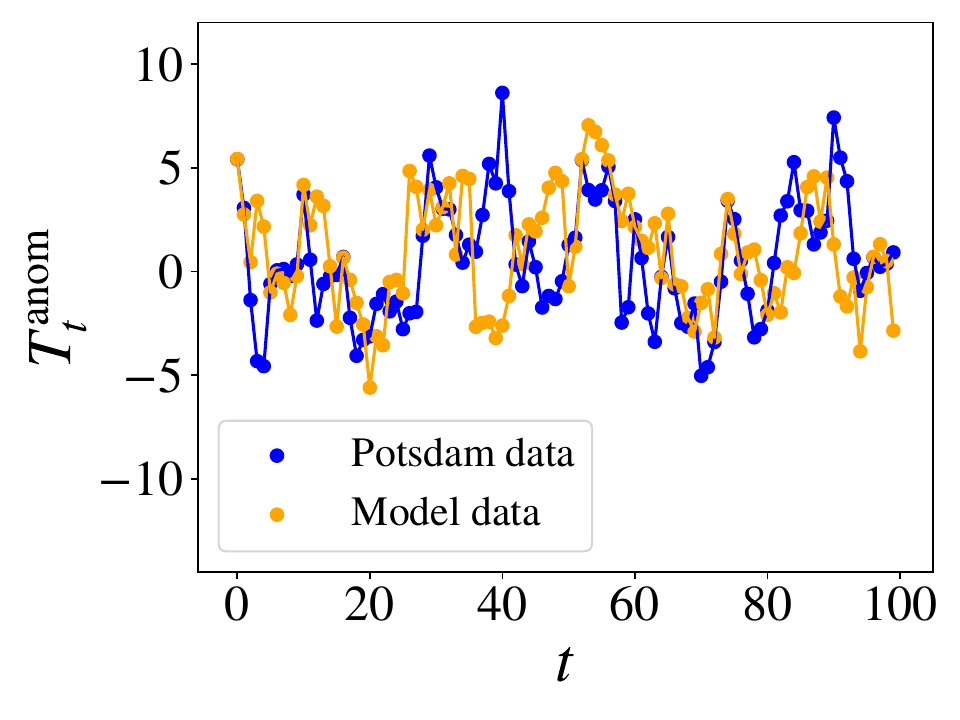}}
		\caption{Comparison of Potsdam daily mean temperature anomalies and model data. For panels~\subref{fig:acf_full_model}, \subref{fig:hist_full_model}, and~\subref{fig:psd_full_model} the model data consists of $100$ samples of the length of the Potsdam data set. The autocorrelation function of the Potsdam data exhibits some small-scale oscillations not explained by our model. The exponential decay of the fractionally differenced Potsdam data is clearly visible, indicating the approximate markovianity of the data. The power spectral density is estimated with a periodogram and Welch's method. The model data shows slightly higher variance than the Potsdam data. The power spectral density of the model agrees well with the Potsdam data apart from a kink at the maximum frequency. Panel~\subref{fig:sample_traj_full_model} shows $100$ data points of the Potsdam daily mean temperature anomalies and one model trajectory conditioned on the past $M=1095$ realizations of the Potsdam daily mean temperature anomalies. Lines between data points are plotted for illustration only.}
		\label{fig:Potsdam_model_comparison}
	\end{figure}

	For the drift and
	diffusion terms, we make a polynomial ansatz of order three and order
	four, respectively. Figure~\ref{fig:KMA} displays the estimated drift
	and diffusion functions for the fractionally differenced
	Potsdam Telegrafenberg daily mean temperature anomalies.
	Kir\'aly and J\'anosi also report nonlinearities for drift and diffusion
	of temperature anomalies for an aggregate of temperature time series of
	20 Hungarian weather stations. \cite[Fig.3, p.4]{Janosi2002}
	Their data shows more pronounced nonlinearities for drift and diffusion
	than the Potsdam temperature anomalies because of more data points
	for large anomalies where nonlinearities are more dominant.

	To ensure numerical stability of the discrete-time Langevin equation
	defined by the estimated drift and diffusion functions, we set
	$\Theta(x_t > x_{\max}) = \Theta(x_{\max})$ and
	$\Theta(x_t < x_{\min}) = \Theta(x_{\min})$.
	We then fractionally integrate a discrete-time Langevin trajectory
	generated with the drift and diffusion parameters obtained.
	Figure~\ref{fig:Potsdam_model_comparison} displays the
	cumulative histograms, autocorrelation functions and power spectral
	densities of the temperature anomalies and model trajectories (see
	Figure~\ref{fig:dfa_anomalies} for the Hurst parameter
	estimation). They are in very good agreement which is also confirmed
	by visual inspection of sample time series
	(cf. Figure~\subref{fig:sample_traj_full_model}).
	
\begin{figure}
		\centering
		\includegraphics[width=.5\textwidth]{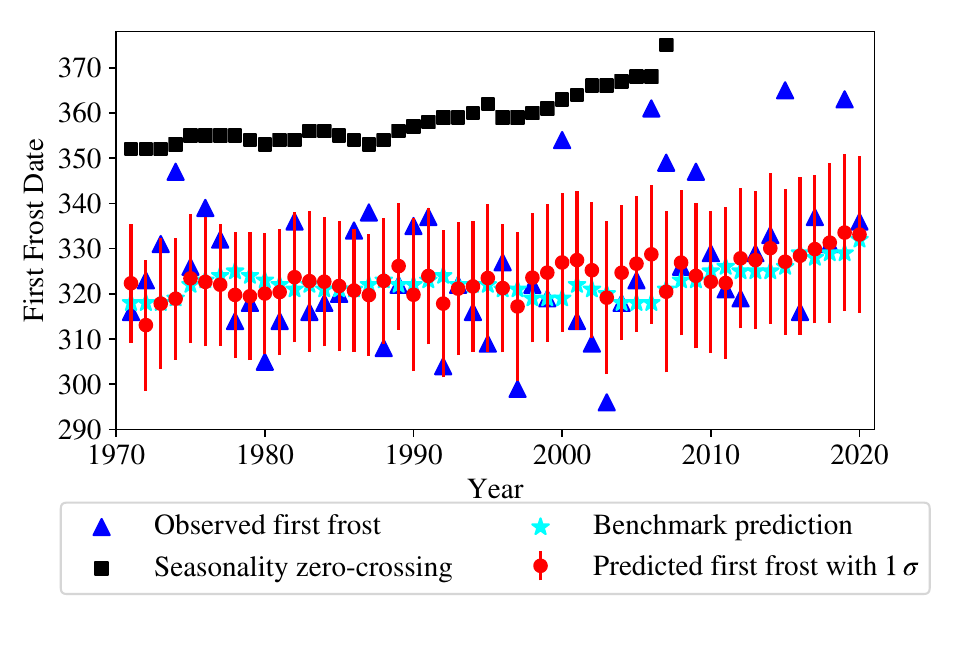}
		\caption{First Frost Prediction Results. Dark blue triangles are the observed first frost dates of the Potsdam Telegrafenberg daily mean temperature data set. Light blue stars indicate the benchmark prediction of the first frost date obtained by fitting a parabola to the previous observed first frost dates since 1893. Black squares are the zero-crossings of the seasonality cycle for years in which they exist. Red dots are the predicted first frost date with one standard deviation of the first frost date distribution. Accuracy of estimators: Prediction: $\text{RMSE}=14.7 \; \mathrm{d}$ and $\text{MAE}=11.4 \; \mathrm{d}$, benchmark prediction: $\text{RMSE}=15.9 \; \mathrm{d}$ and $\text{MAE}=11.9 \; \mathrm{d}$, seasonality: $\text{RMSE}=38.0 \; \mathrm{d}$ and $\text{MAE}=34.8 \; \mathrm{d}$, standard deviation of observed first frost dates: $\sigma=15.6 \; \mathrm{d}$. The first frost prediction performs slightly better than the benchmark prediction.}
		\label{fig:first_frost_results}
	\end{figure}

	The reconstructed process may serve for making predictions. We predict
	the first frost date for the Potsdam Telegrafenberg weather station by
	computing the first passage time distribution of generated process
	trajectories and the zero temperature line for a sample size of fifty years.
	We choose the 23rd of October as the forecast start date. For each sample 
	year, we cut the Potsdam daily mean temperature time series at the 22nd of
	October, resulting in a time series from January
	1st 1893 to the 22nd of October of the sample year. After removal of
	the seasonal cycle, we infer model parameters with our method. Using
	the reconstructed model, we generate $25\times10^3$ trajectories
	using Eq.~\ref{eq:discrete_langevin_model}, setting the fractionally
	differenced temperature on the forecast start date as the initial
	condition. We add the generated trajectory to the fractionally
	differenced temperature anomalies, fractionally integrate the
	concatenated new trajectory, add the seasonal cycle and determine its
	first passage time with the $0\;\si{\celsius}$ temperature line. The mean first
	passage time over the ensemble of $25\times 10^3$ values
	is the predicted first frost date. For a benchmark prediction we fit a parabola to the observed frost dates of the years before the sample year, paralleling the climate change correction, and extrapolate it to the sample year. Figure~\ref{fig:first_frost_results} shows the observed first frost date, the predicted first frost date and its standard deviation, the benchmark prediction and the zero-crossing of the seasonality cycle for the years $1971-2020$.
	The bias of the predicted first frost sample average amounts to
	$-2.9$ days, meaning our prediction only has a marginal bias compared to the average lead time of $32$ days. We use the root-mean-square error (RMSE) and the mean absolute error (MAE) to measure the prediction performance.
	The RMSE of	our prediction is smaller than the variance of the observed first
	frost dates, indicating our prediction narrows the uncertainty of
	the predicted event. RMSE and MAE
	(cf. caption of Figure~\ref{fig:first_frost_results}) show that the
	prediction performs much better than the seasonality but only slightly
	better than the benchmark estimation. We note that the variance of
	the observed first frost date is much larger than the variance of
	the prediction.
	In real weather, the first frost date is impacted by
	many factors, e.g. large-scale weather patterns not captured by the
	local daily mean temperature. Commemorating we solely use a
	one-dimensional time series to predict an event in a high-dimensional
	complex system, we expect better prediction performances for reconstructed
	models in more-dimensional systems. Reconstructing these in multivariate models
	using the method presented in this article is part of future research.
	Additionally, larger values of the memory parameter $d$ would also contribute to larger prediction horizons (cf. Figure~\ref{fig:ar1_arfima}). In meteorology, the first frost date is defined as the first-passage time of the daily minimal	temperature and the zero-degree temperature line whereas we use daily mean temperature data for our analysis. The first frost prediction results for the Potsdam minimal temperature time series are qualitatively identical, but the reconstruction of drift and diffusion is less satisfactory due to their more complex shape.
	
	\section{Conclusion} In this article, we propose a method for the
	reconstruction of one-dimensional nonlinear stochastic processes from
	persistent sparsely sampled time series using fractional
	calculus and discrete-time Langevin equations. The method
	performs well for ARFIMA(1,d,0) and Potsdam daily mean temperature
	data. A first frost prediction for Potsdam daily mean temperature data
	shows predictive power to some extent.
	
	\begin{acknowledgments}
		We thank Philipp G Meyer, Katja Polotzek, Christoph Streissnig, and Benjamin Walter for fruitful discussions and Steffen Peters for IT support.
	\end{acknowledgments}
	
	%

\end{document}